\documentclass[journal]{IEEEtran}

\usepackage{xcolor,soul,framed} %,caption
\usepackage{xcolor}
\colorlet{shadecolor}{yellow}
\usepackage[pdftex]{graphicx}
\graphicspath{{../pdf/}{../jpeg/}}
\DeclareGraphicsExtensions{.pdf,.jpeg,.png}

\usepackage[cmex10]{amsmath}
\usepackage{array}
\usepackage{mdwmath}
\usepackage{mdwtab}
\usepackage{eqparbox}
\usepackage{url}
\usepackage{algorithm,algpseudocode,amsmath}
\usepackage{float}

\usepackage[sorting=none]{biblatex}

\begin{document}
\bstctlcite{IEEEexample:BSTcontrol}
    \title{Towards Energy-Efficient and Low-Latency Voice-Controlled Smart Homes: A Proposal for Offline Speech Recognition and IoT Integration}
    \author{
        \textsuperscript{[1]}~Peng Huang, \textsuperscript{[2]}~Imdad Ullah, \textsuperscript{[3]}~Xiaotong Wei, \\
         \textsuperscript{[4]}~Tariq Ahamed Ahanger, \textsuperscript{[5]}~Najm Hassan
         \textsuperscript{[6]}~Zawar Hussain Shah\\
        {\small
            \textsuperscript{[1, 2, 3]}~School of Computer Science, Faculty of Engineering, The University of Sydney, Sydney NSW 2006, Australia. \\
            \textsuperscript{[4]}~Management Information Systems Department, College of Business Administration, Prince Sattam bin Abdulaziz University, Al-Kharj 16278, Saudi Arabia. \textsuperscript{[5]}~Higher Colleges of Technology, United Arab Emirates (UAE).\\
            \textsuperscript{[6]}~Department of Information Technology, Sydney International School of Technology and Commerce, Sydney NSW 2000, Australia.\\
            \vspace{2mm}
            \textsuperscript{[1, 2, 3]}~phua0038@uni.sydney.edu.au, imdad.ullah@sydney.edu.au, xwei0200@uni.sydney.edu.au, \\
             \textsuperscript{[4]}~t.ahanger@psau.edu.sa,
            \textsuperscript{[5]}~nhassan@hct.ac.ae
            \textsuperscript{[6]}~zawar.s@sistc.nsw.edu.au
        } 
    }

\maketitle

% === ABSTRACT ====================================================================
% =================================================================================
\begin{abstract}
%\boldmath

The smart home systems, based on AI speech recognition and IoT technology, enable people to control devices through verbal commands and make people's lives more efficient.
However, existing AI speech recognition services are primarily deployed on cloud platforms on the Internet.
When users issue a command, speech recognition devices like ``Amazon Echo'' will post a recording through numerous network nodes, reach multiple servers, and then receive responses through the Internet.
This mechanism presents several issues, including unnecessary energy consumption, communication latency, and the risk of a single-point failure.
In this position paper, we propose a smart home concept based on offline speech recognition and IoT technology: 1) integrating offline keyword spotting (KWS) technologies into household appliances with limited resource hardware to enable them to understand user voice commands; 2) designing a local IoT network with decentralized architecture to manage and connect various devices, enhancing the robustness and scalability of the system.
This proposal of a smart home based on offline speech recognition and IoT technology will allow users to use low-latency voice control anywhere in the home without depending on the Internet and provide better scalability and energy sustainability.

\end{abstract}

% === KEYWORDS ====================================================================
% =================================================================================
\begin{IEEEkeywords}
AI, Speech recognition, Voice recognition, IoT, Smart home, Energy.
\end{IEEEkeywords}

% For peer review papers, you can put extra information on the cover
% page as needed:
% \ifCLASSOPTIONpeerreview
% \begin{center} \bfseries EDICS Category: 3-BBND \end{center}
% \fi
%
% For peerreview papers, this IEEEtran command inserts a page break and
% creates the second title. It will be ignored for other modes.
\IEEEpeerreviewmaketitle

% =================================================================================
% =================================================================================
% =================================================================================
\section{Introduction}

% Background
With the rapid development of smart home systems, the intelligence of home environments has improved significantly.
Compared to traditional remote control and automation functions, smart home systems that integrate AI speech recognition and IoT technology enhance their interactivity and connectivity, allowing people to control any device in the system through verbal commands~\cite{8984175}.
In situations where traditional control methods are impractical, such as when users are physically distant from the control panel or struggling to efficiently input complex instructions, voice interaction becomes crucial~\cite{10230286, portet2013design}.
The speech interaction function allows users to use the smart home system more efficiently.
Currently, most AI speech recognition services are predominantly run on cloud platforms, as demonstrated by widely used products like Amazon Alexa, Google Home, and Apple HomeKit.
These technologies have steadily permeated numerous households, demonstrating widespread adoption~\cite{scoop_2024_smart}.

% Research Field
Although speech recognition technology supported by cloud platforms, deployed over the Internet, has been widely integrated into smart home systems, it benefits users' lives.
However, such reliance on cloud processing also introduces problems with interaction latency and energy consumption.
In addition, as more devices are incorporated into the smart home system, the system's robustness and scalability face severe challenges.
This study primarily focuses on the smart home domain, particularly on integrating speech recognition and IoT technology, namely, smart home systems that support speech interaction control and feedback and utilize IoT technology to connect multiple devices.

% Research Gap
Existing solutions typically rely on dedicated devices, such as smart speakers, to capture and record voice commands, send them to cloud-based platforms for recognition, generate feedback and control instructions, and finally relay those instructions to the user’s home devices~\cite{8433167}.
However, this approach presents several challenges:
\textbf{1)} High energy consumption: Each step in the process involves numerous devices, collectively consuming significant energy. It's unnecessary to consume so much power when the user intends to turn on a light in the room.
\textbf{2)} Network dependency: The network quality impacts the latency of interaction between users and servers, which can severely affect user experience in the case of slower Internet speed or when several devices are connected and the network is congested. Hence, the speech interaction features may become unavailable during the network break.
\textbf{3)} Single point of failure: Centralized architecture, which consists of a single speech interaction access node and multiple control nodes, carries the risk of single-point failure. The system will lose speech interaction capability if the access node becomes unavailable.
Therefore, exploring more reliable, higher energy efficiency, and more efficient smart home system solutions is vital.

% Research Objectives & Motivations
Motivated by this, we present a novel proposal for offline speech recognition integrated with the energy-efficient IoT system that provides a low-latency voice-controlled smart home solution. This study aims to
\textbf{1)} integrate offline KWS (keyword spotting) technologies into household appliances with limited resource hardware to enable them to understand user voice commands; 
\textbf{2)} design a local IoT network with decentralized architecture to manage and connect various devices, enhancing the robustness and scalability of the system.

% =================================================================================
% =================================================================================
% =================================================================================
\section{Background}

In this section, we discuss Amazon Alexa as a representative example to introduce the widespread online voice recognition services and networking technologies in smart home systems.

% =================================================================================
\subsection{Online smart home system with voice recognition}

Alexa, provided by Amazon, is a voice service that operates on cloud platforms. It offers voice services through Internet-enabled terminal devices, such as playing news or music and controlling room lighting and temperature~\cite{amazon_alexa_2024}.
Common terminal devices equipped with Alexa include Echo speakers. Additionally, it can be embedded in TVs, wearable devices, and mobile applications for smartphones and tablets ~\cite{alexa_built_in_2024}.

Once deployed, the Echo speaker remains in an idle state until activated into interaction states by the user through a wake-up word like ``Alexa'', tap-to-talk, or push-to-talk methods, unless the microphone has been disabled.
The typical interaction states include:
1) Listening, where the microphone records the user’s voice request and streams it to the Alexa Cloud until the user stops speaking;
2) Thinking, where the Alexa Cloud processes the uploaded request and generates a response to send back to the Echo speaker;
3) Speaking, where the Echo speaker uses text-to-speech (TTS) to play the response to the user's request~\cite{avs_ux_2024}.

The interaction between the Echo speaker and the Alexa Cloud can be viewed as a series of data transmission and processing activities based on an Internet client-server (CS) architecture~\cite{holden2018making}.
Internet-based services have brought convenience to various fields, including e-commerce, streaming services, and mobile applications~\cite{kumar2019review, zebari2019real}.
Despite these benefits, this mechanism still presents limitations in the smart home domain and home usage scenarios.
The instability of Internet connections can impact the delay in the interaction between the home and cloud platform, severely affecting user experience, especially when immediate feedback is desired~\cite{peng2020understanding}.
Voice recognition functionality entirely relies on Internet connectivity and cloud services, rendering it unusable during network outages or cloud infrastructure failures~\cite{Carsen2024, Sabrina2022}.
In rural areas or developing countries lacking quality Internet infrastructure, families cannot access stable Internet connections, which hinders the adoption of smart home systems~\cite{sovacool2020smart}.

% =================================================================================
\subsection{Network connectivity in smart home}

Network connectivity in smart homes allows devices to exchange information with each other. Alexa supports two approaches for connecting with smart home devices, enabling users to control and query the status of smart home devices via voice interactions.

\begin{figure*}
    \centering
    \includegraphics[width=0.77\linewidth]{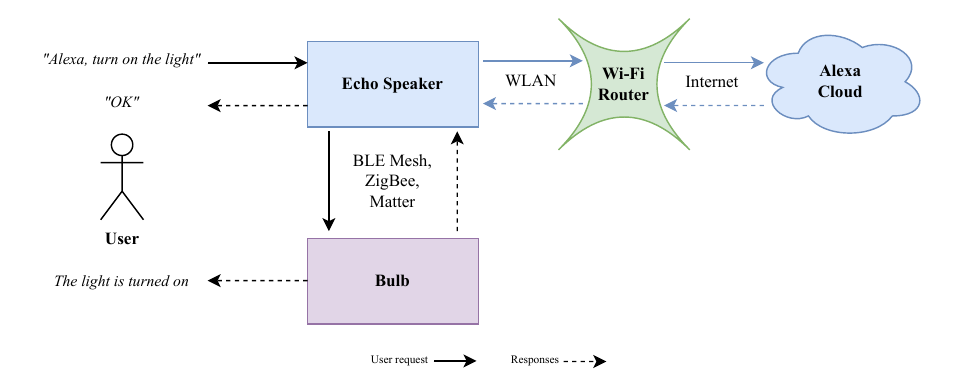}
    \caption{A typical smart home system, where smart home devices connect to the Echo Speaker via local connectivity protocols such as BLE Mesh, ZigBee, and Matter. The user issues a request via the Echo Speaker, which the Alexa Cloud processes to determine intent. The Echo Speaker controls the target device (user receives action response). The Echo Speaker reports device execution results, which the Alexa Cloud processes to generate a response played back through the Echo Speaker (the user receives a voice response). Communication between home devices and the Clouds is facilitated through the Wi-Fi router. Due to the quantity and diversity of elements, the figure does not depict ISP and other relevant cloud infrastructure.}
    \label{fig:alexa_gw}
\end{figure*}

The Echo speaker serves as a ``gateway'' using local connection protocols such as Bluetooth Low-Energy (BLE) Mesh, Matter, and ZigBee IoT technologies, as illustrated in Figure~\ref{fig:alexa_gw}~\cite{alexa_skills_2024}.
BLE Mesh, proposed by The Bluetooth Special Interest Group (SIG), creates a mesh topology facilitated by underlying BLE Scanning and Advertising, using a flooding mechanism for message relaying and forwarding~\cite{baert2018bluetooth}.
Matter, formerly known as Project Connected Home over IP (CHIP) and proposed by the Connectivity Standards Alliance (CSA), aims to provide interoperability between products from different manufacturers, utilizing Wi-Fi and Thread networks at its core~\cite{zegeye2023connected}.
ZigBee, also proposed by CSA earlier, is widely used in monitoring and sensing research fields, achieving efficient message delivery and network management through a routing mechanism, operating under the IEEE 802.15.4 standard~\cite{safaric2006zigbee, rahman2018provisioning}.
Once Alexa understands a user's request, it can quickly control smart home devices using local connection protocols through the Echo speaker~\cite{alexa_skills_2024}.
Different Echo speaker models support different local connection protocols; for instance, the more affordable Echo Pop supports Wi-Fi, BLE Mesh, and Matter protocols, while the more advanced Echo 4th Gen additionally supports ZigBee Hub features~\cite{alexa_devices_2024}.

% Eddie, please clarify the sentence below.
Connecting smart home devices directly to the Internet, such as via Wi-Fi or wired Ethernet.
These smart home devices are directly linked to their respective manufacturer clouds for online control and monitoring, maintained by the device’s manufacturer~\cite{8433167}.
When a user controls smart home devices through an Echo speaker, the voice data containing the user's request is first uploaded to the Alexa Cloud, understood as directives, i.e., the target device and the action to be performed.
Then, the Alexa cloud uses developed smart home skills to invoke APIs from the manufacturer cloud to submit directives.
Subsequently, the manufacturer cloud sends the action to the target device, receives the results from the target device, and responds to the API calls from the Alexa cloud.
Finally, the Alexa cloud converts the response into appropriate feedback text and, through TTS, converts it into an audio file downloaded and played by the Echo speaker, allowing the user to understand the results~\cite{alexa_sm_skills_2024, alexa_sm_options_2024}.

It is important to note that these approaches inherently rely on the Internet, specifically the home local network and the Alexa and manufacturer clouds.
This leads to several limitations:
\textbf{1)} Each step of interaction among these devices and cloud platforms, including the user's home router and modem, community switch, Internet service provider facilities, and cloud platform infrastructure, involves numerous network devices working together, consuming substantial energy. Millions of devices on the market continuously use electricity to maintain connections~\cite{sovacool2020smart}. When the user intends to turn on a light in the room, such extensive energy consumption is unnecessary.
\textbf{2)} All user requests must pass through the Echo speaker and Wi-Fi router, which may face the risk of a single point of failure. The smart home system may face a shutdown when occupied or not functioning.
Overall, while the online voice recognition services and network connectivity represented by Amazon Alexa have brought convenience to global home users, this mechanism still presents problems, including unnecessary energy consumption, interaction delays, and the risk of single-point failure.

\begin{figure*}
    \centering
    \includegraphics[width=0.77\linewidth]{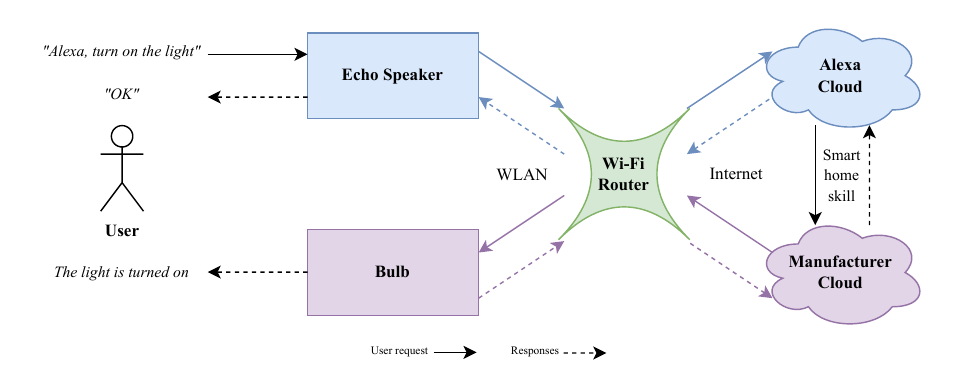}
    \caption{A typical smart home system, where smart home devices are connected to the manufacturer's cloud. The user issues a request via the Echo Speaker, which the Alexa cloud processes to determine the intent. The manufacturer cloud API is then invoked to control the target device (the user receives an action response). The device sends the execution result back to the manufacturer cloud, which then generates a response through the Alexa cloud, and the final feedback is played by the Echo Speaker (the user receives a voice response). The Wi-Fi router must communicate between home devices and the Clouds. Due to the quantity and diversity of elements, the figure does not indicate the ISP and other relevant cloud infrastructure.}
    \label{fig:alexa_manufacturer}
\end{figure*}

% =================================================================================
% =================================================================================
% =================================================================================
\section{Related work}

This section discusses recent research on KWS on resource-constrained platforms and studies on offline smart home systems with voice recognition capabilities.

% =================================================================================
\subsection{KWS on resource-constrained platforms}

KWS is used to identify specific keywords within an audio stream.
Many studies have focused on enhancing the performance and reducing the complexity of KWS to facilitate its integration into small electronic devices such as earphones, smartphones, and smart speakers~\cite{lopez2021deep}.
Traditionally, KWS employed Hidden Markov Models (HMMs) and Gaussian Mixture Models (GMMs) to model keywords and acoustic features~\cite{rohlicek1989continuous}.
Today, neural network-based KWS significantly improves the performance over traditional KWS approaches using architectures such as CNNs, RNNs, and their variants like DS-CNNs (Depthwise Separable Convolutional Neural Networks), LSTMs, GRUs, and CRNNs~\cite{chen2014small, lopez2021deep, sainath2015convolutional, du2018low}.

Researchers have compared the accuracy and memory/computation requirements of different neural network architectures for KWS on resource-constrained microcontrollers.
They quantified pre-trained 32-bit floating-point models into 8-bit fixed-point versions to fit microcontrollers' memory and computational constraints.
These models were validated on Cortex-M7 based STM32F746G MCUs without compromising accuracy.
DS-CNNs have optimized KWS model architectures, achieving better accuracy under the same resource constraints~\cite{zhang2017hello}.

Another study established a CNN model using TensorFlow Lite, consisting of a 2D convolutional layer, a fully connected layer, and a SoftMax layer for output.
The model was converted into C code and deployed on Adafruit EdgeBadge and Arduino Nano 33 BLE Sense development boards for evaluation.
These offline models could recognise 2, 10, and 30 words~\cite{kimvoice}.

Furthermore, researchers proposed OWSNet for performing multiple keyword recognition tasks on Small and medium IoT devices.
They modified the Residual Neural Network~(ResNet) architecture by adopting 1D convolutions along the time axis instead of 2D to reduce the model's parameters, thus speeding up its operation.
OWSNet achieved inference speeds on Raspberry Pi 4 and NVIDIA Jetson Nano platforms that were 1-74 times and 1-12 times faster than other top neural network-based KWS models, respectively, without sacrificing much accuracy~\cite{sudharsan2021owsnet}.

Additionally, researchers introduced a low-power KWS engine featuring an MFCC feature extraction module and an LSTM accelerator.
They trimmed and compressed the LSTM model, reducing its size by 89\% and computation by 76\%, resulting in a 50\% decrease in energy consumption compared to other advanced LSTM KWS hardware.
A simulation using a 40nm CMOS process achieved a power consumption of 2.51 $\mu$W~\cite{chong20212}.

Moreover, researchers combined simulated binary feature extraction with binary neural networks~(BNNs), using an analog front-end instead of digital preprocessing to provide binary time-frequency features, thus alleviating the MCU's computational load.
Compared to previous work, efficiency improved by 4.6$\times$ and accuracy by 1\%.
By adjusting the complexity of the BNN model, a trade-off was achieved where a 2\% drop in accuracy resulted in a 71x energy efficiency improvement~\cite{cerutti2022sub}.

% =================================================================================
\subsection{Offline smart home system with voice recognition}

Offline smart home systems offer voice control services without Internet access.
Researchers introduced an offline smart home system with a voice detection module using a network of cameras with integrated microphones around the house to capture the user's verbal commands.
The data were sent to a Raspberry Pi 3 for voice recognition, action execution, and TTS response.
Experimental results suggested that greater voice intensity and clearer pronouncing could accelerate the system's response time~\cite{ojaghioffline}.

Another approach to integrating voice recognition and IoT technology into smart home systems involved the use of the LD3320 voice module, ESP8266 Wi-Fi module, and Arduino MEGA 2560 MCU, along with multiple peripheral modules for voice controlling indoor lighting, temperature, humidity, and security features.
Researchers used spam keywords to absorb incorrect recognition results, thus enhancing the accuracy of recognition~\cite{wang2019application}.

A control system for indoor household appliances was developed using the EasyVR Commander as a sensor to receive voice commands and then control power switches through an Arduino ATMega328 MCU.
The system supported independent voice recognition for controlling lights, fans, water pumps, and door components and dependent voice recognition for controlling access to safes~\cite{novani2020electrical}.
Similar systems using EasyVR have been developed~\cite{moeid20next}.

HomeIO was proposed as an offline smart home automation system with voice recognition and electricity tracking capabilities.
Researchers deployed a Raspberry Pi 4 with a microphone as the system hub, running voice recognition services and an MQTT Broker.
An ESP32 served as a Smart Plug Socket running an MQTT Client.
A local Wi-Fi Mesh established wireless connections between devices and the hub, enabling offline voice control of household appliance power switches~\cite{irugalbandara2022homeio}.

% =================================================================================
%\subsection{Section Summary {\color{red}(Is this subsection appropriate?)}}

\begin{table*}[h]
    \centering
    \caption{Caption}
    \begin{tabular}{|p{0.6cm}|p{4cm}|p{4cm}|p{4cm}|p{3cm}|}
\hline
\textbf{Year} & \textbf{Title} & \textbf{Proposal} & \textbf{Findings} & \textbf{Limitations} \\
\hline

2017 
& Hello Edge: Keyword Spotting on Microcontrollers ~\cite{zhang2017hello}
& Quantize pre-trained 32-bit floating-point models into 8-bit fixed-point for resource-constrained microcontrollers (Cortex-M7 MCU), comparing accuracy and efficiency across architectures. 
& Models maintained high accuracy on a Cortex-M7 based (STM32F746G) MCU after quantization; DS-CNNs achieved superior accuracy under the same resource constraints
& Potential small accuracy loss due to quantization; limited memory and processing power of MCUs. \\ 
\hline
2017 
& Voice Recognition on Simple Microcontrollers ~\cite{kimvoice}
& Use TensorFlow Lite to run keyword recognition on Arduino microcontrollers. & 2 key words number accuracy achieved 91.7\%, 10 key words accuracy achieved 75.3\%, and 30 key words accuracy achieved 63.5\%. 
& As the vocabulary size increased, the model’s accuracy significantly dropped. A full CMUSphinx system cannot be implemented on Arduino. \\ 
\hline
2021 
& OWSNet: Towards Real-time Offensive Words Spotting Network for Consumer IoT Devices~\cite{sudharsan2021owsnet} & Proposes OWSNet, a lightweight KWS network using temporal convolutions for efficient, low-latency offline detection on mid-range IoT devices, including offensive word recognition.
& OWSNet is faster than other state-of-the-art models, with speed improvements of 1-74 times on Raspberry Pi and 1-12 times on NVIDIA Jetson Nano. 
& Slightly lower accuracy compared to top-performing models. Lacks extensive real-world testing on offensive-language data. \\ 
\hline
2021 & A 2.5 $\mu$W KWS Engine with Pruned LSTM and Embedded MFCC for IoT Applications~\cite{chong20212} & Introduces a power-efficient KWS hardware engine consisting of MFCC-based feature extraction and a LSTM accelerator, optimized via pruning/quantization for deployment on resource-constrained IoT devices.  & 
Achieves 89\% model-size reduction and 76\% fewer computations after LSTM trimming/compression. & Further evaluation needed for noisy or larger-vocabulary scenarios.
 \\ 
\hline
2022 
& Sub-mW Keyword Spotting on an MCU: Analog Binary Feature Extraction and Binary Neural Networks \cite{cerutti2022sub}
& Improve energy efficiency for KWS on MCUs by using BNNs for efficient classification, reducing computational complexity and memory usage. 
& BNN significantly reduced energy consumption during data acquisition by 29$\times$ and saved 71\% of energy output while maintaining accuracy, improving overall system performance. 
& Experimentation and optimization are concentrated on only one hardware platform. \\ 
\hline
\end{tabular}
\end{table*}

\begin{table*}[h]
    \centering
    \caption{Caption}
    \begin{tabular}{|p{0.6cm}|p{4cm}|p{4cm}|p{4cm}|p{3cm}|}
\hline
\textbf{Year} & \textbf{Title} & \textbf{Proposal} & \textbf{Findings} & \textbf{Limitations} \\
\hline

2019
& Application of Speech Recognition Technology in IoT Smart Home~\cite{wang2019application}
& Use LD3320 to capture keywords, execute actions through Arduino, and upload to the cloud via ESP8266.
& Spam keywords can effectively absorb incorrect recognition results.
& Uses centralized architecture. Only one voice entry point.
\\
\hline

2020
& Electrical Household Appliances Control Using Voice Command Based on Microcontroller~\cite{novani2020electrical}
& Use EasyVR to capture keywords and execute actions through Arduino.
& Achieved voice recognition for multiple independent devices.
& Lack of wireless connectivity between devices.
\\
\hline

2020
& Next Generation Home Automation System Based on Voice Recognition~\cite{moeid20next}
& Use EasyVR to capture keywords and execute actions through Arduino.
& An easy-to-learn, cost-effective solution.
& Lack of wireless connectivity between devices.
\\
\hline

2022
& HomeIO: Offline Smart Home Automation System with Automatic Speech Recognition and Household Power Usage Tracking~\cite{irugalbandara2022homeio}
& Use Raspberry Pi with a microphone as a smart hub providing voice recognition functionality.
& Achieved excellent performance in tests compared to other models.
& Uses centralized architecture. Only one voice entry point.
\\
\hline

2023
& Offline Voice Detection in Smart Homes~\cite{ojaghioffline}
& Implement distributed microphones connected via a local network to a Raspberry Pi voice recognition server.
& 1) Higher sound intensity and clarity can speed up response time; 2) System response time is around 1 second, faster than other projects.
& Uses centralized architecture.
\\
\hline

    \end{tabular}
    \label{tab:my_label}
\end{table*}

Recent innovative studies on KWS on resource-constrained platforms have achieved better performance and efficiency than in the past.
With the need for lower power consumption, deploying KWS applications on low-cost hardware has become feasible.
As discussed in Section 2, most voice recognition services are installed on standalone smart speaker hardware, preventing direct voice interaction with smart home devices.
Users cannot control smart home devices through voice if the standalone smart speaker is unavailable.
In homes with multiple rooms or large spaces, it is necessary to appropriately distribute several standalone smart speakers to meet different areas' usage needs.
Therefore, Section 4 proposes integrating KWS directly into household appliance hardware, allowing users to interact directly with appliances via voice, reducing interaction delays and eliminating reliance on Internet connectivity.
Users only need to purchase household appliances without additional equipment and network services to enjoy built-in voice control features.

Research on offline smart home systems with voice recognition often adopts a centralized architecture, using a local home server to run voice recognition services and manage devices through a local network.
This poses the risk of single-point failure; if the server becomes unavailable, the smart home system may cease functioning.
Thus, our proposal in Section 4 adopts a decentralized architecture, including application and network layers.
This architecture does not require a server and allows peer-to-peer communication between devices.
If a single device loses power, only the functions it is responsible for are affected, and users can still control other devices through voice, keeping the smart home system operational.

% =================================================================================
% =================================================================================
% =================================================================================
\section{Proposed Solution}

This section presents the proposed solution for integrating offline speech recognition and IoT technologies in a smart home system.
Initially, we designed a four-layer system architecture. We encapsulated five types of components commonly found in household appliances, along with two typical approaches for integrating KWS units into these appliances.
Subsequently, we outlined a decentralized network in a smart home context with diverse IoT technologies. We described three typical types of voice interactions between users and devices in offline smart home systems.

% =================================================================================
\subsection{Architecture overview}

\begin{figure*}
    \centering
    \includegraphics[width=0.85\linewidth]{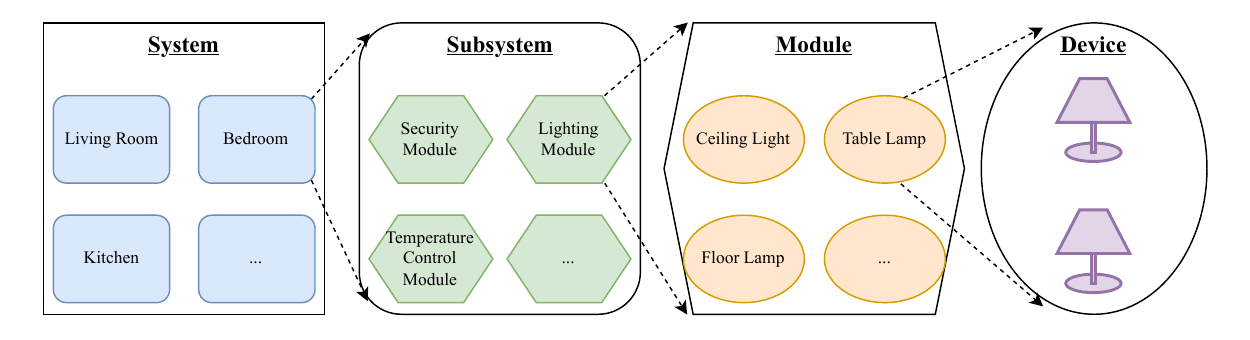}
    \caption{The architecture of the smart home system can be divided into four layers: System Layer, Subsystem Layer, Module Layer, and Device Layer.}
    \label{fig:architecture}
\end{figure*}

To better organize and manage the various functionalities and spaces within the smart home system, we have designed a layered architecture, as depicted in Figure~\ref{fig:architecture}.
This architecture consists of conceptual layers—System, Subsystem, and Module—and a physical layer, Device.
Each layer focuses on different aspects: the System layer on entire household spaces, the Subsystem layer on specific rooms or areas, the Module layer on functions or services, and the Device layer on specific household appliances.
Below is an explanation of each layer in our proposed architecture.

\textbf{System layer} is the highest level of our proposed smart home architecture, organizing the entire household space and defining the overall structure and scope boundaries.
The system includes multiple subsystems, each serving different parts of the household space.
For instance, a smart home system designed for an apartment might include subsystems for the living room, bedroom, kitchen, bathroom, and balcony.
Household appliances within the system achieve connectivity through integrated IoT technology, allowing devices from the same or different subsystems to exchange information and facilitate collaboration in specific scenarios.
The system does not require Internet access to enable speech recognition capabilities, as we will implement offline speech recognition services within the devices in our system, specifically by integrating a KWS unit.
This allows users to manage and control the household appliances in the system through offline speech interactions.

\textbf{Subsystem layer} contains multiple subsystems, each representing different areas within the household spaces.
An area can be a specific purpose room or a shared space.
Depending on the purpose of the area, each subsystem is composed of different modules to fulfil the required functionalities of that area.
For example, subsystems exist for the living room, bedroom, and balcony.
The living room subsystem may include multimedia and lighting modules;
the bedroom subsystem may consist of sleep assist and lighting modules;
and the balcony subsystem may consist of lighting and security modules.

\textbf{Module layer} focuses on specific functions or services, such as lighting, temperature control, security, etc.
A module may consist of one or more devices with a common purpose.
For example, a temperature control module might only include an air conditioning unit, while a lighting module could contain multiple devices such as ceiling lights, table lamps, and floor lamps.
Different modules cooperate to provide a complete smart home experience for users.

\textbf{Device layer} contains specific household appliance devices, such as table lamps and ceiling lights (belonging to the lighting module), heaters (belonging to the temperature control module), cameras (belonging to the security module), etc.
Devices are the physical implementation of module functionalities; these hardware entities directly engage in user interactions within the smart home, executing specific tasks and functions.

With its clearly defined layers, the proposed architecture brings the advantage of efficient management.
The system can quickly aggregate and control devices within specific modules or subsystems, enhancing overall response speed and control efficiency and better coordinating interactions and functional integration between different layers.
The modular design also provides system flexibility and maintainability, allowing users to easily add, remove, or modify devices without affecting other system parts.

% =================================================================================
\subsection{Household Appliances}

\begin{figure*}
    \centering
    \includegraphics[width=0.85\linewidth]{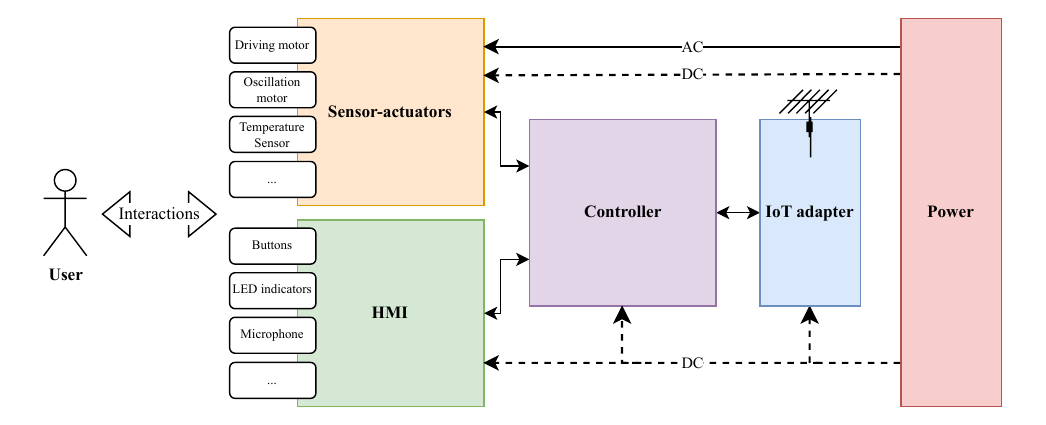}
    \caption{The internal elements of household appliances can be categorized into five components based on their purposes.}
    \label{fig:appliances}
\end{figure*}

A comprehensive smart home system comprises various household appliances serving different purposes.
We categorize the internal elements of these appliances into five components based on their functionalities, as illustrated in Figure~\ref{fig:appliances}.
Below is an explanation of each component.

\textbf{Power} component supplies the necessary voltage and current for the appliance's operation, such as power adapters and batteries.
The power component can be divided into multiple channels depending on the usage.
The power channel used for driving actuators may be either Alternating Current (AC) or Direct Current (DC), such as motors and heating devices, which often require more significant power.
In contrast, power channels for controllers, Human-Machine Interactions (HMI), and network connectivity utilize lower power DC supplies.
Notably, in our proposed solution, integrating a KWS unit within the household appliance may alter the DC power requirements.
Therefore, it is necessary to evaluate and adapt the power components to meet these new requirements.

\textbf{Sensor-actuators} implement the appliance's functionality, including peripherals for output (e.g., motors) and sensors for input (e.g., temperature sensors in an air conditioner).

\textbf{Controller}, typically a microcontroller unit~(MCU), is the primary computational unit that coordinates the functions of various elements within the appliance and maintains device operation.
Controllers generally operate continuously from the moment the device is powered, as they need to capture user button presses at any time, parse network messages, read from sensors if there are changes in the environment, and update the operation modes of actuators.

\textbf{Human-Machine Interaction (HMI)} units designed to receive or sense user intentions and output feedback or alert the user with specific information.
Examples include buttons for input and LED indicators for output.

\textbf{IoT adapter} units for communicating with other devices in the network system, such as BLE chips, ZigBee chips, Wi-Fi chips, etc.

In contrast to traditional household appliances, we will integrate a KWS unit into the HMI components to enable voice interaction capabilities. This lets users directly control any appliance using voice commands through their built-in capabilities, eliminating the need for a separate smart speaker.

% =================================================================================
\subsection{Keyword spotting unit} 

\begin{figure*}
    \centering
    \includegraphics[width=0.95\linewidth]{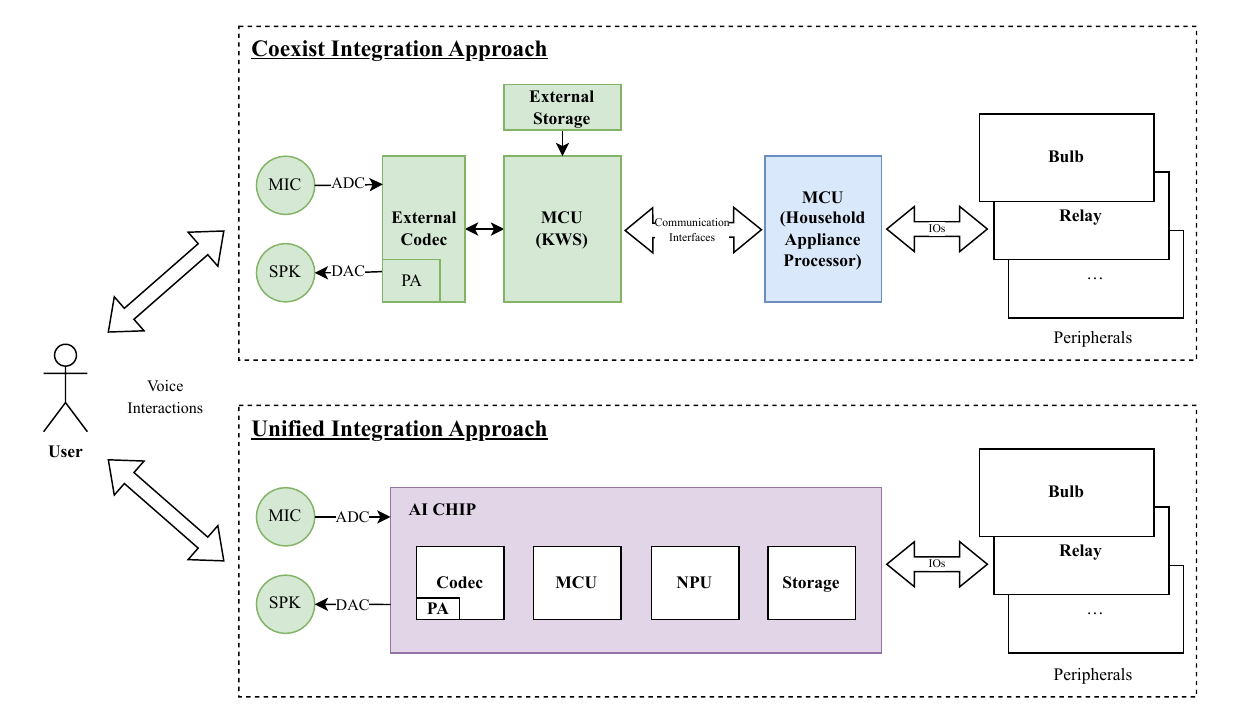}
    \caption{Two approaches to integrating KWS units into household appliances: The Coexist Integration Approach establishes a communication interface between the original appliance MCU and the KWS units, thereby enabling the appliance to acquire KWS capabilities. The Unified Integration Approach utilizes an advanced AI chip to replace the original appliance MCU role and acquire KWS capabilities.}
    \label{fig:KWS}
\end{figure*}

We propose two typical approaches to integrating KWS units into household appliances, as shown in Figure~\ref{fig:KWS}.

\textbf{Coexist Integration Approach:}
The KWS algorithm is deployed on an independent MCU, which inputs microphone signals and outputs speaker signals by connecting to an external Audio Encoder \& Decoder Hardware~(Codec), with standard communication interfaces between the Codec and the MCU being Inter-IC Sound Bus~(I$^2$S) and Inter-Integrated Circuit~(I$^2$C)~\cite{UM11732, UM10204}.
The external storage, pre-programmed with response sound files, is accessed by the KWS MCU through interfaces such as Serial Peripheral Interface~(SPI)~\cite{ADG1412YCPZREEL7}.
In some traditional home appliances, where an internal MCU already implements the business logic and peripheral control functions, the KWS speech interaction functionality can be achieved by connecting to the communication interface of the MCU running the KWS algorithm.
Standard communication interfaces include Universal Asynchronous Receiver-Transmitter~(UART), I$^2$C, SPI, etc~\cite{nanda2016universal, UM10204, ADG1412YCPZREEL7}.
These interfaces can serve as buses, meaning they do not significantly increase the occupation of Input/output Interfaces (IOs)~\cite{6220723}.
For hardware platforms with limited IO resources, single-wire communication protocols can be implemented through one IO~\cite{6220723, ninikrishna20161}.

Algorithm~\ref{alg:1} describes the pseudocode for implementing voice-controlled appliances through the Coexist Integration Approach.
Initially, the user utters a wake-up word to activate the device into a wake state, during which it can receive command words.
The variable $t$ represents the countdown timer to exit the wake state; once it expires, the device will ignore command words to prevent unintended actions.
$T$ is the default duration of the wake window, for example, 10 seconds.
The variable $cont$ is used to allow one or multiple command words during a single wake period.

The advantage of the Coexist Integration Approach is the rapid integration of KWS voice interaction capabilities.
It adds a KWS unit to existing appliances without requiring significant modifications.
The main task involves adding a communication interface between the existing MCU and the KWS MCU, modifying the firmware of the existing MCU to receive messages from the KWS MCU and performing corresponding actions.

\begin{algorithm}[h]
  \caption{Coexist Integration Approach}
    \textbf{Input:} verbal command \\
    \textbf{Output:} voice response, action response
  \begin{algorithmic}
    \\\hrulefill
    \State \textbf{Initialization}
  \end{algorithmic}
  \begin{algorithmic}[1]
    \State Establish communication between the KWS MCU and the appliance MCU.
    \State The Appliance MCU initializes all appliance peripherals and functions.
    \State The KWS MCU initializes connections with external storage and codec.
  \end{algorithmic}
  \begin{algorithmic}
    \\\hrulefill
    \State \textbf{Main Loop of KWS MCU}
  \end{algorithmic}
  \begin{algorithmic}[1]
    \State Continuously capture audio signals and detect keywords.
    \If{a wake-up word is spotted}
        \State $t \gets T$
        \State Play wake-up voice response.
        \State Notify appliance MCU to wake up.
    \EndIf
    \If{a command word is spotted \textbf{and} $t > 0$}
        \State Notify appliance MCU to perform action.
        \If{$cont$ is True}
            \State $t \gets T$
        \Else
            \State $t \gets 0$
        \EndIf
    \EndIf
    \If{$t > 0$}
        \State Update $t$ until it reaches $0$
    \EndIf
    \If{receive notification from appliance MCU}
        \State Play specified voice response.
    \EndIf
  \end{algorithmic}
  \begin{algorithmic}
    \\\hrulefill
    \State \textbf{Main Loop of Appliance MCU}
  \end{algorithmic}
  \begin{algorithmic}[1]
    \State Maintain appliance functionalities.
    \If{receive wake-up notification from KWS MCU}
        \State Display wake-up indication (optional).
    \ElsIf{receive action notification from KWS MCU}
        \If{action is valid}
            \State Perform action.
            \State Notify KWS MCU to play successful audio.
        \Else
            \State Notify KWS MCU to play failed audio.
        \EndIf
    \EndIf
  \end{algorithmic}
  \label{alg:1}
\end{algorithm}

\textbf{Unified Integration Approach:}
This approach uses a specialized AI chip developed for KWS applications, which internally integrates a Neural Processing Unit~(NPU), Codec, MCU, and storage units.
A representative product is the Voitist 811 and its series from Intengine, featuring a high-speed NPU accelerator and hardware-accelerated functionalities for Fast Fourier Transform (FFT), Mel-Frequency Cepstral Coefficients (MFCC), Voice Activity Detection (VAD), Acoustic Echo Cancellation (AEC), and other audio signal processing tasks, providing high-speed, low-energy KWS/NLP inference computations.
It allows for embedding up to 300 keywords, providing feedback within 0.2 seconds once a verbal command is completed.
At an approximate cost of \$1.50, it offers 23 GPIOs and multiple groups of SPI, I2C, UART, PWM, ADC, etc. peripherals~\cite{VOI811}.
This single AI chip can potentially implement all functionalities required by the Coexist Integration Approach.
The Unified Integration Approach offers a simple yet flexible hardware design solution.
Developers can select the optimal AI chip model based on peripheral IOs requirements and the number of KWS words~\cite{VOI811}.
These AI chips also offer adaptability and scalability.
Firmware developed using C/C++, the most prevalent programming language on embedded platforms, forms the basis of these chips.
The accompanying Software Development Kit~(SDK) typically operates with a Real-Time Operating System~(RTOS), allowing developers to integrate existing code by creating new threads.
This facilitates easy integration into existing MCU functions within appliances, such as reading sensor signals and controlling actuators.
These AI chips support horizontal scaling, for instance, by paralleling multiple AI chips to increase the number of recognizable keywords for offline speech recognition.
Additionally, they enable vertical scaling by attaching external flash storage and pseudo-static random-access memory (PSRAM) chips.

Algorithm~\ref{alg:2} outlines the pseudocode for implementing voice-controlled appliances using the Unified Integration Approach. Compared to the Coexist Integration Approach, it eliminates the communication interface between the two chips, thereby reducing the program's complexity.

The Unified Integration Approach simplifies hardware and software design, resulting in highly cohesive systems that can be deployed on smaller hardware with lower power consumption.
Since all software runs on a single chip, it features a higher degree of integration and lower latency.
It uses a single AI chip with a KWS application to replace the work of the original MCU, with main tasks including modifying hardware and porting existing code to the new platform.

\begin{algorithm}
  \caption{Unified Integration Approach}
    \textbf{Input:} verbal command \\
    \textbf{Output:} voice response, action response

  \begin{algorithmic}
    \\\hrulefill
    \State \textbf{Initialization}
  \end{algorithmic}

  \begin{algorithmic}[1]
    \State Initialize internal units of AI chip, appliance peripherals, and functions.
  \end{algorithmic}

  \begin{algorithmic}
    \\\hrulefill
    \State \textbf{Main Loop of AI chip}
  \end{algorithmic}

  \begin{algorithmic}[1]
    \State Maintain appliance functionalities.
    \State Continuously capture audio signals and detect keywords.
    \If{a wake-up word is spotted}
        \State $t \gets T$
        \State Play wake-up voice response.
        \State Display wake-up indication (optional).
    \EndIf
    \If{a command word is spotted \textbf{and} $t > 0$}
        \If{action is valid}
            \State Perform action.
            \State Play successful voice response.
        \Else
            \State Play failed voice response.
        \EndIf
        \If{$cont$ is True}
            \State $t \gets T$
        \Else
            \State $t \gets 0$
        \EndIf
    \EndIf
    \If{$t > 0$}
        \State Update $t$ until it reaches $0$
    \EndIf
  \end{algorithmic}
  \label{alg:2}
\end{algorithm}

% =================================================================================
\subsection{Networks}

Smart home systems comprise multiple devices connected to the local network as nodes, thereby gaining the ability to communicate and collaborate with other devices.
The network's fundamental function is to transmit messages from one node to others, with message types categorized into:
Unicast (one-to-one), such as in a temperature control module where an indoor humidity sensor can read the current environmental humidity and send a message through the local network to activate a dehumidifier if it's too humid or a humidifier if it's too dry, allowing the indoor humidity to be maintained within an appropriate range;
multicast (one-to-many), such as in a lighting module where a brightness sensor can detect the current ambient light and notify multiple ceiling lights to adjust the brightness based on the local time;
and broadcast (one-to-all), such as when a user leaves home, broadcasting a message to all devices to switch to away mode, enhancing security and reducing energy consumption.

Besides basic communication capabilities, a robust local network also requires a self-healing mechanism.
When changes occur in the network, such as when users add new devices or remove existing ones, or if devices unexpectedly lose power, the network needs to automatically adjust its topology based on the remaining devices' quantity, roles, and locations to adapt to these changes.

To achieve decentralization and mitigate the risk of a single-point failure, we divide our network solution into two parts:
\textbf{1)} The lower-level fundamental communication part, responsible for physical, link, and network layer transactions, encapsulating messages from upper-level applications into packets for transmission over the air or cable, and receiving packets from the air or cable that are destined for the current node and updating the upper-level application.
\textbf{2)} The upper-level application communication part, responsible for implementing smart home business logic and product functionalities, encryption, delivery of specific control commands, and content.

The lower-level fundamental communication part is based on mesh networking, with widely researched and used technologies including BLE Mesh, ZigBee, and Wi-Fi Mesh~\cite{9353075, 7757102, 8920920, MUHENDRA2017332}.
Compared to star and tree networks, mesh networks offer higher reliability and fault tolerance.
Common message propagation mechanisms can be divided into two types:
\textbf{1)} For flooding mechanisms, such as BLE Mesh, the failure of a single node or path does not cause a network outage, as data can be rerouted through other paths around the faulty node~\cite{9035389, darroudi2017bluetooth}.
\textbf{2)} For routing mechanisms, such as ZigBee, self-organizing networks can dynamically generate their network topology to adapt to changes~\cite{5555486}.
Mesh networks also provide better scalability, not limited by the capacity and coverage of a single central node.
New nodes can interact with one or more existing nearby nodes, easily extending the network's coverage by adding new nodes~\cite{darroudi2017bluetooth}.
These mesh network technologies have distinct features, giving them advantages in different aspects~\cite{4460126, 8757472}.
Considering the diversity of household appliances, no single solution can meet all devices' needs.
A robust smart home system network should combine the strengths of various IoT technologies, using appropriate technologies for specific devices.
Using BLE Mesh or ZigBee for battery-powered devices can help save more power, thus extending battery life.
For devices powered by alternating current, Wi-Fi Mesh can provide relatively stronger network performance~\cite{4460126}.
It is possible to integrate multiple IoT technologies into a single device, such as the ESP32 series system-on-chip~(SoC) products that support both Wi-Fi and BLE, giving such devices the potential to act as gateways~\cite{8101926, 8695968}.

\begin{figure}
    \centering
    \includegraphics[width=1\linewidth]{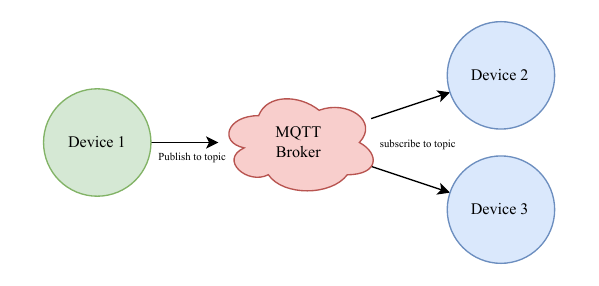}
    \caption{The subscription-publish model of the MQTT protocol involves devices 1, 2, and 3, all connecting to the MQTT Broker. Devices 2 and 3 have subscribed to a specific topic. Once device 1 publishes a new message on this topic, the Broker pushes the message to all devices subscribed to the topic, resulting in devices 2 and 3 receiving the message from device 1.}
    \label{fig:mqtt}
\end{figure}

\begin{figure}
    \centering
    \includegraphics[width=1\linewidth]{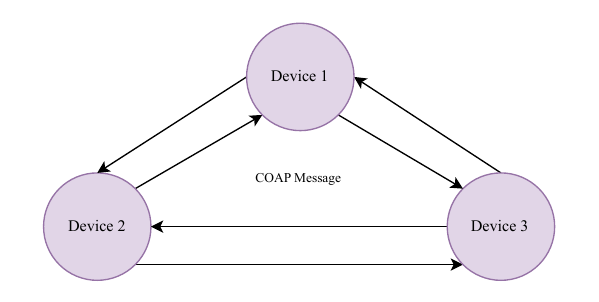}
    \caption{In the client-server model of the CoAP protocol, devices 1, 2, and 3 form a decentralized structure. Each device simultaneously deploys both client and server roles, enabling it to initiate requests to and respond to requests from other devices. This facilitates bidirectional communication between devices without the need for a centralized Broker.}
    \label{fig:coap}
\end{figure}

The upper-level application communication part also employs a decentralized design.
Widely used IoT communication protocols include Message Queuing Telemetry Transport (MQTT) and Constrained Application Protocol (CoAP)~\cite{8088251}.
MQTT uses a publish-subscribe model and requires a centralized server as a Broker to relay messages, as shown in Figure~\ref{fig:mqtt}~\cite{9247996}.
In contrast, CoAP uses a server/client model to run both client and server on a device, enabling direct communication between devices, as shown in Figure~\ref{fig:coap}~\cite{ansari2018Internet}.
Moreover, CoAP uses the UDP protocol, reducing overhead and handshake delays compared to MQTT's TCP, further saving hardware resources and power consumption~\cite{8088251, 8985850}.
Given that CoAP does not require a centralized server as a Broker compared to MQTT, we currently consider CoAP more suitable for distributed network applications, especially in offline smart home systems~\cite{8088251}.
Because CoAP is based on the UDP protocol, it can be easily deployed on devices integrated with Wi-Fi technology, with less application in BLE Mesh and ZigBee networks.
BLE Mesh currently does not support the TCP/IP protocol stack. ZigBee has a derivative version, ZigBee IP, which supports IPv6~\cite{6673344}.

To simultaneously accommodate multiple networks in a smart home system, gateways can act as bridges between different IoT technologies.
For example, when you leave home, and a battery-powered passive infrared sensor~(PIR) sensor no longer detects activity signals, it initiates a ZigBee network message to notify the TV to enter sleep mode, but the TV only has Wi-Fi capability and not ZigBee.
Therefore, gateway devices compatible with ZigBee and Wi-Fi networks can help achieve this goal, forwarding messages from the ZigBee network to the Wi-Fi network, allowing the TV to receive the event notification to enter sleep mode, thereby saving users on electricity costs.

% =================================================================================
\subsection{Operations}

As proposed earlier, integrating KWS units into various household appliances and creating a decentralized smart home network can meet the diverse usage needs of smart home scenarios.
We have defined three types of voice interactions between users and devices.

\begin{figure}[H]
    \centering
    \includegraphics[width=0.66\linewidth]{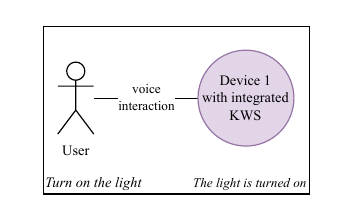}
    \caption{Direct device interaction, where the user interacts directly with the target device integrated with a Keyword Spotting (KWS) unit. The target device understands, executes, and provides feedback. This interaction is independent of network activity.}
    \label{fig:direct-device}
\end{figure}

\textbf{Direct device interaction} refers to users directly interacting with the target device they wish to operate through verbal commands.
The device understands, executes, and provides feedback on the user's verbal commands without relying on network activity, as shown in Figure~\ref{fig:direct-device}.
Even if there is only one device in the smart home system, as long as it integrates a KWS unit, the user can interact with this sole device using verbal commands without the need for the Internet or other external resources.
For example, a user on the balcony says the verbal command "Turn on the light," which is heard and understood by a ceiling light on the balcony, and then the ceiling light executes the action of turning on the light.
Due to the limited distance the sound travels, the user needs to speak the verbal command near the target device; if the user is too far from the target device, the sound may not be clearly perceived.
The main advantage of direct device interaction is that the device can immediately provide feedback on the action with minimal response delay in its interaction.

\begin{figure}[H]
    \centering
    \includegraphics[width=1\linewidth]{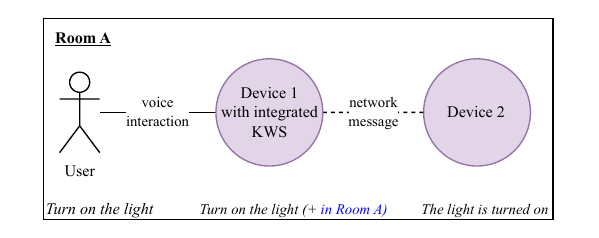}
    \caption{Cross-device Interaction (within a subsystem), where the user interacts with a target device through another device integrated with a Keyword Spotting (KWS) unit. The user and the two devices are located in Room A. Listening device 1 automatically populates current room attributes into the user's intent and notifies the target device of actions through network messages.}
    \label{fig:Cross-device-within-subsystems}
\end{figure}

\begin{figure}[H]
    \centering
    \includegraphics[width=1\linewidth]{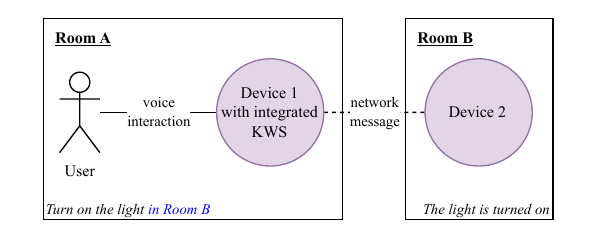}
    \caption{Cross-device Interaction (between subsystems), where the user interacts with a target device in Room B through a device integrated with a Keyword Spotting (KWS) unit in Room A. The user explicitly includes the room's name where the target device is located in their verbal command.}
    \label{fig:Cross-device-between-subsystems}
\end{figure}

\textbf{Cross-device interaction within a subsystem} refers to a user controlling devices within the current room via verbal commands.
Suppose the target device has not integrated a KWS unit or cannot adequately hear the user's voice. In that case, this task is achieved through a device with better sound reception quality that has integrated a KWS unit within the same room, as shown in Figure~\ref{fig:Cross-device-within-subsystems}.
In this operation, the user does not need to include the current location word in their verbal command because the listening device and target device are within the same subsystem, allowing them to share each other's location information.
Once the user issues a verbal command without specifying the target location, the listening device that hears the command will automatically fill in the current room attribute to the user's intent and create a network message, enabling other devices within the subsystem to understand the cross-device request.
For example, a user commands a washing machine located on the balcony, which lacks an integrated KWS unit, by saying "start washing" to a ceiling light equipped with a KWS unit on the same balcony.
The ceiling light detects the command, assigns the balcony attribute, and transmits the message across the local network.
Consequently, this prompts the balcony's washing machine to commence operation, rather than the washing machine situated in the bathroom.

\textbf{Cross-device interaction between subsystems} refers to a user interacting vocally with devices in another room from their current location, as shown in Figure~\ref{fig:Cross-device-between-subsystems}.
In this case, users need to explicitly include the name of the target room in their verbal commands to ensure the message is correctly relayed to the corresponding subsystem.
For example, if users in the living room want to turn off the bedroom light, they can say the verbal command, "Turn off the bedroom light." This command is heard by a device with an integrated KWS unit in the living room and relayed through the smart home system network to the light devices in the bedroom to execute.
Message transmission may involve multiple hops between the source and target device nodes, thereby introducing more latency than the first two interaction types.
However, it still occurs within the local network, has a shorter physical transmission distance, and is unaffected by Internet quality.

% =================================================================================
% =================================================================================
% =================================================================================
\section{Discussion}

The proposal to integrate offline KWS technology with a decentralized local IoT network enhances smart home systems' efficiency, robustness, and scalability.
By integrating KWS units into household appliances, low-latency voice control is achieved without relying on an Internet connection. This not only reduces energy consumption but also enhances user experience. The design of a decentralized local IoT mesh network improves the system's scalability and robustness and reduces the risk of a single point of failure.
The market's predominant smart home systems with voice recognition focus on cloud-based voice recognition services, such as Amazon Alexa. While some researchers have proposed offline smart home systems with voice recognition, most of these studies have utilized a centralized architecture that depends on a single device for voice recognition services. Our research offers a solution that operates independently of the Internet and a decentralized local network, filling existing research gaps.
This proposed smart home system has broad application prospects, especially in areas where Internet connections are unstable or unavailable. By processing voice commands offline, the smart home system's response speed and reliability are improved while reducing dependence on network infrastructure.
%{\color{blue}
The system proposed in this study merits further exploration under various conditions, such as increased node counts, high traffic loads, and conflicting commands.
The continuous addition of nodes places ongoing pressure on network capacity. Although the mesh network design allows for a large number of nodes compared to star or tree networks, which no longer depend on a root node's capacity, the nodes that act as relays—receiving and transmitting data and thereby extending the local network's coverage—will face increased data transmission tasks. Thus, the configuration of these relays could impact the system's scalability.
High traffic conditions further challenge the performance of the local network as increased queue processing and forwarding wait times cause communication delays. Once traffic exceeds the processing capabilities of the devices or bandwidth limitations, packet loss occurs. Although packets can be resent, this also means delays. Optimizing routing mechanisms and using directed flooding can mitigate these effects.
Conflicting commands, either from multiple users or from misrecognition by the KWS, demand that a target device perform opposite actions, such as turning on and off in a short span, which could damage the device or lead to poor user experience. Incorporating decision-making algorithms within the system can resolve these conflicts.%}
The limitations of this study include the lack of large-scale practical application testing. The performance of offline KWS technology in different environments and the security and management of decentralized networks require further research.

% =================================================================================
% =================================================================================
% =================================================================================
\section{Conclusion}

This study aims to propose a smart home system based on offline voice recognition and IoT technology to address the unnecessary energy consumption, interaction delays, and single-point failure risks associated with existing cloud-based voice recognition services.
Our solution integrates KWS units into household appliances and designs a decentralized local IoT network to achieve low-latency voice control, enhancing system robustness and scalability.
This system allows users to interact vocally with household appliances directly, without relying on any Internet connection or external resources, reducing unnecessary energy consumption and contributing to sustainable energy development.
This research introduces new ideas and solutions to the smart home technology field, fills a research gap in existing studies, and provides important references for future technological developments.
Future research can continue to explore the performance of KWS on resource-constrained platforms and the porting of more powerful large language models (LLMs) to these chips to explore more application scenarios and conduct large-scale practical application tests. Additionally, the security and management strategies of decentralized networks could be investigated to ensure the secure and reliable operation of the system.

% =================================================================================
% =================================================================================
% =================================================================================
\printbibliography

@INPROCEEDINGS{8984175,
  author={Wang, Peng and Lu, Xiang and Sun, Hongyu and Lv, Wenhong},
  booktitle={2019 IEEE 3rd Advanced Information Management, Communicates, Electronic and Automation Control Conference (IMCEC)}, 
  title={Application of speech recognition technology in IoT smart home}, 
  year={2019},
  volume={},
  number={},
  pages={1264-1267},
  doi={10.1109/IMCEC46724.2019.8984175}}

@INPROCEEDINGS{8433167,
  author={Lei, Xinyu and Tu, Guan-Hua and Liu, Alex X. and Li, Chi-Yu and Xie, Tian},
  booktitle={2018 IEEE Conference on Communications and Network Security (CNS)}, 
  title={The Insecurity of Home Digital Voice Assistants - Vulnerabilities, Attacks and Countermeasures}, 
  year={2018},
  volume={},
  number={},
  pages={1-9},
  doi={10.1109/CNS.2018.8433167}}

@article{portet2013design,
  title={Design and evaluation of a smart home voice interface for the elderly: acceptability and objection aspects},
  author={Portet, Fran{\c{c}}ois and Vacher, Michel and Golanski, Caroline and Roux, Camille and Meillon, Brigitte},
  journal={Personal and Ubiquitous Computing},
  volume={17},
  pages={127--144},
  year={2013},
  publisher={Springer}
}

@ARTICLE{10230286,
  author={Norda, Marvin and Engel, Christoph and Rennies, Jan and Appell, Jens-E. and Lange, Sven Carsten and Hahn, Axel},
  journal={IEEE Transactions on Automation Science and Engineering}, 
  title={Evaluating the Efficiency of Voice Control as Human Machine Interface in Production}, 
  year={2023},
  volume={},
  number={},
  pages={1-12},
  doi={10.1109/TASE.2023.3302951}}

@misc{scoop_2024_smart,
  title        = {Smart Speaker Statistics: Best Voice Control Technology},
  author       = {Tajammul Pangarkar},
  year         = 2024,
  month        = Jun,
  howpublished = {Web},
  url          = {https://scoop.market.us/smart-speaker-statistics/},
  note         = {Accessed: 2024-06-20}
}

@misc{amazon_alexa_2024,
  title        = {Amazon Alexa Voice AI | Alexa Developer Official Site},
  author       = {{Amazon}},
  year         = 2024,
  month        = Jun,
  howpublished = {Web},
  url          = {https://developer.amazon.com/en-US/alexa},
  note         = {Accessed: 2024-06-23}
}

@misc{alexa_built_in_2024,
  title        = {Alexa Built-in Devices Official Site: Create Voice-First Devices},
  author       = {{Amazon}},
  year         = 2024,
  month        = Jun,
  howpublished = {Web},
  url          = {https://developer.amazon.com/en-US/alexa/devices/alexa-built-in},
  note         = {Accessed: 2024-06-23}
}

@misc{avs_ux_2024,
  title        = {AVS UX Attention System |Alexa Voice Service},
  author       = {{Amazon}},
  year         = 2024,
  month        = Jun,
  howpublished = {Web},
  url          = {https://developer.amazon.com/en-US/docs/alexa/alexa-voice-service/ux-design-attention.html},
  note         = {Accessed: 2024-06-23}
}

@mastersthesis{holden2018making,
  title={Making Your Devices Speak. Integration between Amazon Alexa and the Managed IoT Cloud},
  author={Holden, Thomas},
  year={2018},
  school={UiT Norges arktiske universitet}
}

@article{kumar2019review,
  title={A Review on Client-Server based applications and research opportunity},
  author={Kumar, Santosh},
  journal={International Journal of Recent Scientific Research},
  volume={10},
  number={7},
  pages={33857--3386},
  year={2019}
}

@inproceedings{zebari2019real,
  title={Real time video streaming from multi-source using client-server for video distribution},
  author={Zebari, Ibrahim MI and Zeebaree, Subhi RM and Yasin, Hajar Maseeh},
  booktitle={2019 4th Scientific International Conference Najaf (SICN)},
  pages={109--114},
  year={2019},
  organization={IEEE}
}

@inproceedings{peng2020understanding,
  title={Understanding User Perceptions of Robot's Delay, Voice Quality-Speed Trade-off and GUI during Conversation},
  author={Peng, Zhenhui and Mo, Kaixiang and Zhu, Xiaogang and Chen, Junlin and Chen, Zhijun and Xu, Qian and Ma, Xiaojuan},
  booktitle={Extended Abstracts of the 2020 CHI Conference on Human Factors in Computing Systems},
  pages={1--8},
  year={2020}
}

@misc{Carsen2024,
  title        = {Amazon Alexa down updates — Users complain voice assistant connectivity and control are not working as outage underway | The US Sun},
  author       = {{Carsen Holaday}},
  year         = 2024,
  month        = Jun,
  howpublished = {Web},
  url          = {https://www.the-sun.com/tech/11807249/amazon-alexa-down-updates-not-working/},
  note         = {Accessed: 2024-06-25}
}

@misc{Sabrina2022,
  title        = {Amazon's Alexa 'down for thousands of customers' | Metro News},
  author       = {{Sabrina Johnson}},
  year         = 2022,
  month        = Apr,
  howpublished = {Web},
  url          = {https://metro.co.uk/2022/04/16/amazons-alexa-down-for-thousands-of-customers-16479166/},
  note         = {Accessed: 2024-06-25}
}

@article{sovacool2020smart,
  title={Smart home technologies in Europe: A critical review of concepts, benefits, risks and policies},
  author={Sovacool, Benjamin K and Del Rio, Dylan D Furszyfer},
  journal={Renewable and sustainable energy reviews},
  volume={120},
  pages={109663},
  year={2020},
  publisher={Elsevier}
}

@misc{alexa_skills_2024,
  title        = {Local Connectivity Protocol Options for Smart Home Devices | Alexa Skills Kit},
  author       = {{Amazon}},
  year         = 2024,
  month        = Jun,
  howpublished = {Web},
  url          = {https://developer.amazon.com/en-US/docs/alexa/smarthome/wwa-connection-options.html},
  note         = {Accessed: 2024-06-23}
}

@article{baert2018bluetooth,
  title={The Bluetooth mesh standard: An overview and experimental evaluation},
  author={Baert, Mathias and Rossey, Jen and Shahid, Adnan and Hoebeke, Jeroen},
  journal={Sensors},
  volume={18},
  number={8},
  pages={2409},
  year={2018},
  publisher={Mdpi}
}

@inproceedings{zegeye2023connected,
  title={Connected smart home over matter protocol},
  author={Zegeye, Wondimu and Jemal, Ahamed and Kornegay, Kevin},
  booktitle={2023 IEEE International Conference on Consumer Electronics (ICCE)},
  pages={1--7},
  year={2023},
  organization={IEEE}
}

@inproceedings{safaric2006zigbee,
  title={ZigBee wireless standard},
  author={Safaric, Stanislav and Malaric, Kresimir},
  booktitle={Proceedings ELMAR 2006},
  pages={259--262},
  year={2006},
  organization={IEEE}
}

@inproceedings{rahman2018provisioning,
  title={Provisioning technical interoperability within ZigBee and BLE in IoT environment},
  author={Rahman, Taibur and Chakraborty, Swarnendu Kumar},
  booktitle={2018 2nd International Conference on Electronics, Materials Engineering \& Nano-Technology (IEMENTech)},
  pages={1--4},
  year={2018},
  organization={IEEE}
}

@misc{alexa_devices_2024,
  title        = {Amazon.com.au: Echo \& Alexa devices: Amazon Devices \& Accessories},
  author       = {{Amazon}},
  year         = 2024,
  month        = Jun,
  howpublished = {Web},
  url          = {https://www.amazon.com.au/Alexa-Amazon-Echo-Smart-Speakers/b?ie=UTF8&node=5425434051},
  note         = {Accessed: 2024-06-25}
}

@misc{alexa_sm_skills_2024,
  title        = {Understand Smart Home Skills | Alexa Skills Kit},
  author       = {{Amazon}},
  year         = 2024,
  month        = Jun,
  howpublished = {Web},
  url          = {https://developer.amazon.com/en-US/docs/alexa/smarthome/understand-the-smart-home-skill-api.html},
  note         = {Accessed: 2024-06-25}
}

@misc{alexa_sm_options_2024,
  title        = {Smart Home Development Options | Alexa Skills Kit},
  author       = {{Amazon}},
  year         = 2024,
  month        = Jun,
  howpublished = {Web},
  url          = {https://developer.amazon.com/en-US/docs/alexa/smarthome/development-options.html},
  note         = {Accessed: 2024-06-25}
}

@article{lopez2021deep,
  title={Deep spoken keyword spotting: An overview},
  author={L{\'o}pez-Espejo, Iv{\'a}n and Tan, Zheng-Hua and Hansen, John HL and Jensen, Jesper},
  journal={IEEE Access},
  volume={10},
  pages={4169--4199},
  year={2021},
  publisher={IEEE}
}

@inproceedings{rohlicek1989continuous,
  title={Continuous hidden Markov modeling for speaker-independent word spotting},
  author={Rohlicek, Jan Robin and Russell, William and Roukos, Salim and Gish, Herbert},
  booktitle={International Conference on Acoustics, Speech, and Signal Processing,},
  pages={627--630},
  year={1989},
  organization={IEEE}
}

@inproceedings{chen2014small,
  title={Small-footprint keyword spotting using deep neural networks},
  author={Chen, Guoguo and Parada, Carolina and Heigold, Georg},
  booktitle={2014 IEEE international conference on acoustics, speech and signal processing (ICASSP)},
  pages={4087--4091},
  year={2014},
  organization={IEEE}
}

@inproceedings{sainath2015convolutional,
  title={Convolutional neural networks for small-footprint keyword spotting.},
  author={Sainath, Tara N and Parada, Carolina},
  booktitle={Interspeech},
  pages={1478--1482},
  year={2015}
}

@inproceedings{du2018low,
  title={Low-latency convolutional recurrent neural network for keyword spotting},
  author={Du, Hu and Li, Ruohan and Kim, Donggyun and Hirota, Kaoru and Dai, Yaping},
  booktitle={2018 Joint 10th International Conference on Soft Computing and Intelligent Systems (SCIS) and 19th International Symposium on Advanced Intelligent Systems (ISIS)},
  pages={802--807},
  year={2018},
  organization={IEEE}
}

@article{zhang2017hello,
  title={Hello edge: Keyword spotting on microcontrollers},
  author={Zhang, Yundong and Suda, Naveen and Lai, Liangzhen and Chandra, Vikas},
  journal={arXiv preprint arXiv:1711.07128},
  year={2017}
}

@article{kimvoice,
  title={Voice Recognition on Simple Microcontrollers},
  author={Kim, Callie Y and Chou, Diana and Liu, Geng}
}

@inproceedings{sudharsan2021owsnet,
  title={Owsnet: Towards real-time offensive words spotting network for consumer iot devices},
  author={Sudharsan, Bharath and Malik, Sweta and Corcoran, Peter and Patel, Pankesh and Breslin, John G and Ali, Muhammad Intizar},
  booktitle={2021 IEEE 7th World Forum on Internet of Things (WF-IoT)},
  pages={83--88},
  year={2021},
  organization={IEEE}
}

@article{chong20212,
  title={A 2.5 $\mu$W KWS engine with pruned LSTM and embedded MFCC for IoT applications},
  author={Chong, Yi Sheng and Goh, Wang Ling and Nambiar, Vishnu P and Do, Anh-Tuan},
  journal={IEEE Transactions on Circuits and Systems II: Express Briefs},
  volume={69},
  number={3},
  pages={1662--1666},
  year={2021},
  publisher={IEEE}
}

@article{cerutti2022sub,
  title={Sub-mW keyword spotting on an MCU: Analog binary feature extraction and binary neural networks},
  author={Cerutti, Gianmarco and Cavigelli, Lukas and Andri, Renzo and Magno, Michele and Farella, Elisabetta and Benini, Luca},
  journal={IEEE Transactions on Circuits and Systems I: Regular Papers},
  volume={69},
  number={5},
  pages={2002--2012},
  year={2022},
  publisher={IEEE}
}

@inproceedings{ojaghioffline,
  title={Offline voice detection in smart homes},
  author={Ojaghi, Sina and Ghafourian, Javid and Maleki, Pouria and Hedayatnia, Atefeh},
  year={2023},
  booktitle={The 4th International Conference on Electrical Engineering, Computer, Mechanics and Artificial Intelligence},
}

@inproceedings{wang2019application,
  title={Application of speech recognition technology in IoT smart home},
  author={Wang, Peng and Lu, Xiang and Sun, Hongyu and Lv, Wenhong},
  booktitle={2019 IEEE 3rd Advanced Information Management, Communicates, Electronic and Automation Control Conference (IMCEC)},
  pages={1264--1267},
  year={2019},
  organization={IEEE}
}

@inproceedings{novani2020electrical,
  title={Electrical household appliances control using voice command based on microcontroller},
  author={Novani, Nefy Puteri and Hersyah, Mohammad Hafiz and Hamdanu, Ryon},
  booktitle={2020 International Conference on Information Technology Systems and Innovation (ICITSI)},
  pages={288--293},
  year={2020},
  organization={IEEE}
}

@article{moeid20next,
  title={Next generation home automation system based on voice recognition},
  author={Moeid, ME and Hana, HS and Amer, RZ},
  journal={ICEMIS},
  volume={20},
  pages={1--7}
}

@inproceedings{irugalbandara2022homeio,
  title={HomeIO: Offline smart home automation system with automatic speech recognition and household power usage tracking},
  author={Irugalbandara, IB Chandra and Naseem, ASM and Perera, MSH and Logeeshan, Velmanickam},
  booktitle={2022 IEEE World AI IoT Congress (AIIoT)},
  pages={571--577},
  year={2022},
  organization={IEEE}
}

@manual{UM11732,
  title = {I2S bus specification},
  author = {{NXP Semiconductors}},
  year = {2022},
  url = {https://www.nxp.com/docs/en/user-manual/UM11732.pdf},
  note = {Accessed: 2024-07-05}
}

@manual{UM10204,
  title = {I2C-bus specification and user manual},
  author = {{NXP Semiconductors}},
  year = {2021},
  url = {https://www.nxp.com/docs/en/user-guide/UM10204.pdf},
  note = {Accessed: 2024-07-05}
}

@manual{ADG1412YCPZREEL7,
  title = {Introduction to SPI Interface},
  author = {{Analog Devices}},
  year = {2018},
  url = {https://www.allelcoelec.in/datasheets.b8/ADG1412YCPZ-REEL7.pdf},
  note = {Accessed: 2024-07-05}
}

@inproceedings{nanda2016universal,
  title={Universal asynchronous receiver and transmitter (uart)},
  author={Nanda, Umakanta and Pattnaik, Sushant Kumar},
  booktitle={2016 3rd international conference on advanced computing and communication systems (ICACCS)},
  volume={1},
  pages={1--5},
  year={2016},
  organization={IEEE}
}

@INPROCEEDINGS{6220723,
  author={Rezaei, Hamid Fahim and Kruger, Anton},
  booktitle={2012 IEEE International Conference on Electro/Information Technology}, 
  title={Wired-FM, a novel distributed digital single-wire field bus}, 
  year={2012},
  volume={},
  number={},
  pages={1-4},
  doi={10.1109/EIT.2012.6220723}
}

@inproceedings{ninikrishna20161,
  title={1-Wire Communication Protocol for Debugging Modem Chipsets},
  author={Ninikrishna, T and Sahu, Soni and Thalaiappan, Rathina Balan and Haricharan, Siri and Manjunath, RK},
  booktitle={Emerging Research in Computing, Information, Communication and Applications: ERCICA 2015, Volume 2},
  pages={535--548},
  year={2016},
  organization={Springer}
}

@manual{VOI811,
  title = {VOI811 Datasheet V2.3},
  author = {{intenginetech}},
  year = {2023},
  url = {https://www.intenginetech.com/},
  note = {Accessed: 2024-07-08}
}

@INPROCEEDINGS{9353075,
  author={Tran, Quoc Thi and Tran, Dai Duong and Doan, Duy and Nguyen, Minh Son},
  booktitle={2020 International Conference on Advanced Computing and Applications (ACOMP)}, 
  title={An Approach of BLE Mesh Network For Smart Home Application}, 
  year={2020},
  volume={},
  number={},
  pages={170-174},
  doi={10.1109/ACOMP50827.2020.00034}
}

@INPROCEEDINGS{7757102,
  author={Zhihua, Su},
  booktitle={2016 International Conference on Robots \& Intelligent System (ICRIS)}, 
  title={Design of Smart Home System Based on ZigBee}, 
  year={2016},
  volume={},
  number={},
  pages={167-170},
  doi={10.1109/ICRIS.2016.35}
}

@INPROCEEDINGS{8920920,
  author={Son, Vo Que and Khoa, Nguyen Tho Anh},
  booktitle={2019 International Symposium on Electrical and Electronics Engineering (ISEE)}, 
  title={Evaluation of Full-Mesh Networks for Smart Home Applications}, 
  year={2019},
  volume={},
  number={},
  pages={73-78},
  doi={10.1109/ISEE2.2019.8920920}}

@article{MUHENDRA2017332,
    title = {Development of WiFi Mesh Infrastructure for Internet of Things Applications},
    journal = {Procedia Engineering},
    volume = {170},
    pages = {332-337},
    year = {2017},
    note = {Engineering Physics International Conference 2016 – EPIC 2016},
    issn = {1877-7058},
    doi = {https://doi.org/10.1016/j.proeng.2017.03.045},
    url = {https://www.sciencedirect.com/science/article/pii/S1877705817311566},
    author = {Rifki Muhendra and Aditya Rinaldi and Maman Budiman and  Khairurrijal},
}

@ARTICLE{9035389,
  author={Hernández-Solana, ángela and Pérez-Díaz-De-Cerio, David and García-Lozano, Mario and Bardají, Antonio Valdovinos and Valenzuela, José-Luis},
  journal={IEEE Access}, 
  title={Bluetooth Mesh Analysis, Issues, and Challenges}, 
  year={2020},
  volume={8},
  number={},
  pages={53784-53800},
  doi={10.1109/ACCESS.2020.2980795}
}

@article{darroudi2017bluetooth,
  title={Bluetooth low energy mesh networks: A survey},
  author={Darroudi, Seyed Mahdi and Gomez, Carles},
  journal={Sensors},
  volume={17},
  number={7},
  pages={1467},
  year={2017},
  publisher={MDPI}
}

@INPROCEEDINGS{5555486,
  author={Li, Jianpo and Zhu, Xuning and Tang, Ning and Sui, Jisheng},
  booktitle={2010 2nd International Conference on Signal Processing Systems}, 
  title={Study on ZigBee network architecture and routing algorithm}, 
  year={2010},
  volume={2},
  number={},
  pages={V2-389-V2-393},
  doi={10.1109/ICSPS.2010.5555486}
}

@INPROCEEDINGS{4460126,
  author={Lee, Jin-Shyan and Su, Yu-Wei and Shen, Chung-Chou},
  booktitle={IECON 2007 - 33rd Annual Conference of the IEEE Industrial Electronics Society}, 
  title={A Comparative Study of Wireless Protocols: Bluetooth, UWB, ZigBee, and Wi-Fi}, 
  year={2007},
  volume={},
  number={},
  pages={46-51},
  doi={10.1109/IECON.2007.4460126}
}

@INPROCEEDINGS{8757472,
  author={Danbatta, Salim Jibrin and Varol, Asaf},
  booktitle={2019 7th International Symposium on Digital Forensics and Security (ISDFS)}, 
  title={Comparison of Zigbee, Z-Wave, Wi-Fi, and Bluetooth Wireless Technologies Used in Home Automation}, 
  year={2019},
  volume={},
  number={},
  pages={1-5},
  doi={10.1109/ISDFS.2019.8757472}
}

@INPROCEEDINGS{8101926,
  author={Maier, Alexander and Sharp, Andrew and Vagapov, Yuriy},
  booktitle={2017 Internet Technologies and Applications (ITA)}, 
  title={Comparative analysis and practical implementation of the ESP32 microcontroller module for the internet of things}, 
  year={2017},
  volume={},
  number={},
  pages={143-148},
  doi={10.1109/ITECHA.2017.8101926}
}

@INPROCEEDINGS{8695968,
  author={Khanchuea, Kanitkorn and Siripokarpirom, Rawat},
  booktitle={2019 10th International Conference of Information and Communication Technology for Embedded Systems (IC-ICTES)}, 
  title={A Multi-Protocol IoT Gateway and WiFi/BLE Sensor Nodes for Smart Home and Building Automation: Design and Implementation}, 
  year={2019},
  volume={},
  number={},
  pages={1-6},
  doi={10.1109/ICTEmSys.2019.8695968}
}

@INPROCEEDINGS{8088251,
  author={Naik, Nitin},
  booktitle={2017 IEEE International Systems Engineering Symposium (ISSE)}, 
  title={Choice of effective messaging protocols for IoT systems: MQTT, CoAP, AMQP and HTTP}, 
  year={2017},
  volume={},
  number={},
  pages={1-7},
  doi={10.1109/SysEng.2017.8088251}}

@ARTICLE{9247996,
  author={Mishra, Biswajeeban and Kertesz, Attila},
  journal={IEEE Access}, 
  title={The Use of MQTT in M2M and IoT Systems: A Survey}, 
  year={2020},
  volume={8},
  number={},
  pages={201071-201086},
  doi={10.1109/ACCESS.2020.3035849}
}

@article{ansari2018internet,
  title={Internet of things (iot) protocols: a brief exploration of mqtt and coap},
  author={Ansari, Danish Bilal and Rehman, Atteeq-Ur and Ali, Rizwan},
  journal={International Journal of Computer Applications},
  volume={179},
  number={27},
  pages={9--14},
  year={2018}
}

@INPROCEEDINGS{8985850,
  author={Prayogo, Sandy Suryo and Mukhlis, Yulisdin and Yakti, Bayu Kumoro},
  booktitle={2019 Fourth International Conference on Informatics and Computing (ICIC)}, 
  title={The Use and Performance of MQTT and CoAP as Internet of Things Application Protocol using NodeMCU ESP8266}, 
  year={2019},
  volume={},
  number={},
  pages={1-5},
  doi={10.1109/ICIC47613.2019.8985850}
}

@INPROCEEDINGS{6673344,
  author={Franceschinis, Mirko and Pastrone, Claudio and Spirito, Maurizio A. and Borean, Claudio},
  booktitle={2013 IEEE 9th International Conference on Wireless and Mobile Computing, Networking and Communications (WiMob)}, 
  title={On the performance of ZigBee Pro and ZigBee IP in IEEE 802.15.4 networks}, 
  year={2013},
  volume={},
  number={},
  pages={83-88},
  doi={10.1109/WiMOB.2013.6673344}
}

\vfill

% Can be used to pull up biographies so that the bottom of the last one
% is flush with the other column.
%\enlargethispage{-5in}

% that's all folks
\end{document}